%% file: MAIN_OH_in_Smith_Cloud.tex
\documentclass[manuscript]{aastex631}
\graphicspath{{./}{figures/}}

\begin{document}

\newcommand{\cmm}{\ifmmode{\rm cm^{-2}}\else{$\rm cm^{-2}$}\fi}
\newcommand{\hi}{\ifmmode{\rm HI}\else{H\/{\sc i}}\fi} 
\newcommand{\oh}{\ifmmode{\rm OH}\else{O\/{\sc H}}\fi} 
\newcommand{\glon}{\ifmmode{\ell}\else{$\ell$}\fi} 
\newcommand{\glat}{\ifmmode{b}\else{$b$}\fi}
\newcommand{\vlsr}{\ifmmode{V_\mathrm{LSR}}\else{$V_\mathrm{LSR}$}\fi}
\newcommand{\vwind}{\ifmmode{V_\mathrm{w}}\else{$V_\mathrm{w}$}\fi} 
\newcommand{\dg}{\ifmmode{^\circ}\else{$^\circ$}\fi} 
\newcommand {\kms}{\ifmmode{\rm km \, s^{-1}}\else{$\rm km \, s^{-1}$}\fi} 
\newcommand {\mo}{${\rm M}_\odot$}
\newcommand {\Ro}{${\rm R}_\odot$}
\newcommand {\moyr}{\,{\rm M_\odot\,\rm yr}^{-1}}
\newcommand{\nhi}{\ifmmode{\rm N_{HI}}\else{$\rm N_{HI}$}\fi}
\newcommand{\noh}{\ifmmode{\rm N_{OH}}\else{$\rm N_{OH}$}\fi}
\newcommand{\nh}{\ifmmode{\rm N_{H}}\else{$\rm N_{H}$}\fi}

\newcommand{\edt}[1]{{\color{red}#1}}





\shorttitle{\oh\  in the Smith Cloud}
\shortauthors{Minter et al.}

\title{Limits on the OH Molecule in the Smith High Velocity Cloud}

\correspondingauthor{A.H. Minter}
\email{tminter@nrao.edu}

\author[0000-0002-6555-312X]{Anthony H. Minter}
\affiliation{Green Bank Observatory, Green Bank, WV 24944, USA}

\author[0000-0002-6050-2008]{Felix J.\ Lockman}https://www.overleaf.com/project/63e8fa7b858beeec75fd8fdc
\affiliation{Green Bank Observatory, Green Bank, WV 24944, USA}

\author[0000-0002-3814-9666]{S. A.\ Balashev}
\affiliation{Independent Researcher}

\author[0000-0003-4019-0673]{H. Alyson Ford}
\affiliation{Steward Observatory and Department of Astronomy, University of Arizona, 933 N. Cherry Ave., Tucson, AZ 85721}

\begin{abstract}

We have used the Green Bank Telescope (GBT) to search for the \oh\ molecule  at several locations in the Smith Cloud, one of the most prominent of the high-velocity clouds that surround the Milky Way.  
Five positions with a high \hi\ column density were selected as targets for individual pointings, along with a  square degree around a molecular cloud detected with the Planck telescope near the tip of the Smith Cloud.
Gas in the Galactic disk with similar values of \nhi\ has detectable \oh\ emission. 
Although we found \oh\  at velocities consistent with the foreground Aquila molecular cloud, nothing was found at the velocity of the Smith Cloud to an rms level of 0.7 mK (T$_b$) in a 1 \kms\ channel.  
{\bf The three positions that give the strictest limits on \oh\ are analyzed in detail.
Their combined data } imply a $5\sigma$ limit on ${\rm N(H_2)/\nhi  \leq 0.03 }$ scaled by a factor dependent on  the \oh\ excitation temperature and background continuum $T_{ex}/(T_{ex}-T_{bg})$.
There is no evidence for far-infrared emission from dust within the Smith Cloud. 
These results are consistent with expectations for a low-metallicity 
diffuse cloud exposed to the radiation field of the Galactic halo {\bf rather than a product of a galactic fountain.}

\end{abstract}


\section{Introduction}

\input{Introduction.tex}

\begin{figure}
	\centering
	\includegraphics[width=0.5\textwidth]{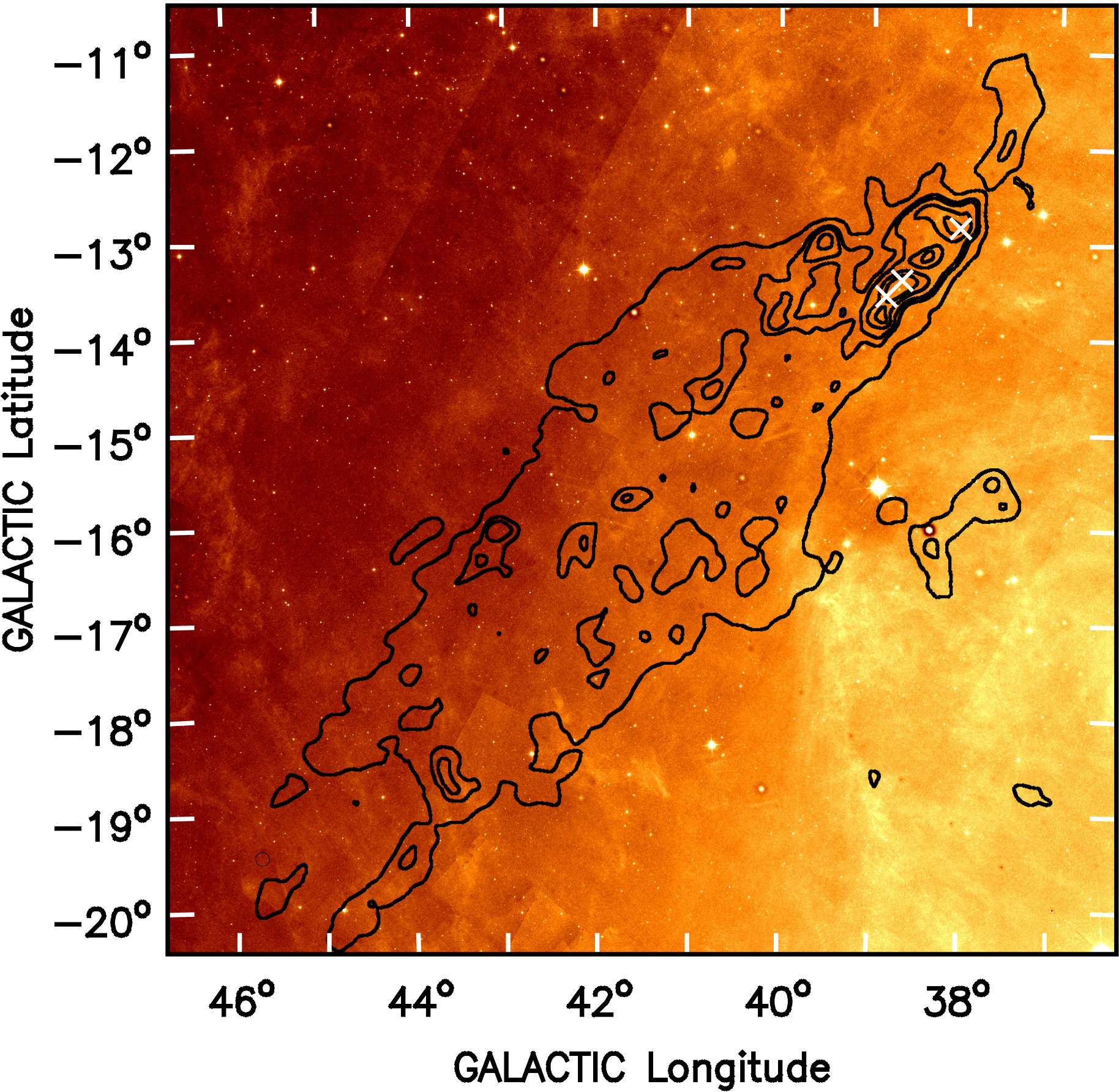}
	\caption{
        \hi\ column density 
		in contours plotted atop WISE 22$\mu$m image that shows emission from the dust along the line of sight.
	White crosses show the locations of the individual \oh\ pointings described in Tables 2 and 3.
  HI contours are drawn at $0.5$, $1.4$, $1.8$, $2.9$, and $3.8 \times 10^{20}$ \cmm. 
  The WISE 22 $\mu$m image has arbitrary units as obtained from NASA's HEASARC SkyView.}
		\label{fig:OH_pointings}
\end{figure}

\section{Observations}
\label{sec:Observations}
Observations of the 18 cm lines of \oh\ and the 21cm line of \hi\ were made with the GBT \citep{2009Prestage} using the L band receiver and the VEGAS spectrometer.  
The GBT has an angular resolution of $7.5\arcmin$ and $9.1\arcmin$ at the frequencies of the \oh\ and \hi\ lines, respectively, {\bf and the main beam efficiency is 0.83 for both species.
 Observational parameters are summarized in Table 1, where $V_{LSR}$ is the velocity range covered, and ${\delta v}$ is the channel spacing.  
Col.~6 gives the typical rms noise in a channel and col.~5 the typical integration time spent at each of the three most important positions. 
}

\begin{deluxetable}{cccccc}
	\tablecolumns{6}
	\tablecaption{GBT Observational Summary} 
\label{tab:ObsSummary}
\tablehead{\colhead{Species}  & \colhead{HPBW} & 
\colhead{${ V_{LSR}}$ range} &
\colhead{$\delta v$} & \colhead{$t_{int}$\tablenotemark{a}} & 
\colhead{$\sigma T_b $\tablenotemark{a}} 
\\
 & \colhead{($\arcmin$)} & \colhead{(\kms)} & \colhead{(\kms)}  & \colhead{(min)}  & \colhead{(mK)}    \\
\colhead{(1)} &  \colhead{(2)} & \colhead{(3)} & \colhead{(4)}  & \colhead{(5)} & \colhead{(6)}
}
\startdata
	\hline\hline\noalign{\vspace{5pt}}	
			\noalign{\smallskip}
\hi & 9.1 & $\pm400$ & 1.03 & 0.3 & 67 \\
\oh & 7.5 & $-585 \rightarrow +800$  & 1.10 & 2700 & 0.7 \\
\noalign{\vspace{5pt}}
\enddata
\tablenotetext{a}{Typical values}
\end{deluxetable}

\subsection{HI Observations}

\input{HI_observations.tex}

\input{Smith_HI_tab.tex}

\subsection{OH Observations}
\label{sec:OHObservations}

Spectra in the four 18 cm transitions of \oh\ were taken toward the three directions having some of the largest values of \nhi\ in the Smith Cloud, as well as a one square-degree area around the tip of the cloud.  
Data were acquired using  in-band frequency switching.  
The sub-reflector was moved around the focus position by one-eighth of the 18cm wavelength 
to reduce the impact of standing waves in the spectra.
The intrinsic velocity resolution of the \oh\ spectra is 1.10 \kms, sufficient to resolve any expected emission.  For consistency we report noise levels normalized to a 1 \kms\ velocity channel.

All of the OH data were flagged for undesired signals -- radio frequency interference (RFI) -- that appears at  frequencies near the OH lines, especially the 1612 MHz line.  
First a derivative of each spectrum was determined by subtracting a copy of the spectrum shifted by one channel.  
A robust fit to the mean of the derivative was performed and points more than $5\sigma$ from the mean were identified and flagged as strong RFI.  
The second step involved a robust fit of a polynomial to the band-pass of each spectrum.  
Again, data deviating more than $5\sigma$ from the fit were flagged as RFI.  
The final step was to split the data into two sets by the time of observation and by polarization.
As thermal main-line OH emission from interstellar clouds is not expected to vary in time or to be linearly polarized, any signal that exhibited either behavior was assumed to be spurious.
This procedure identified RFI  that could have otherwise been mistaken for a real signal.

No \oh\ emission was detected that could be attributed to the Smith Cloud.
The only measurable \oh\ lines are at a low \vlsr\ and arise in the  foreground  Aquila molecular cloud \citep{Dame2001}.
The \oh\ observations are summarized in Table 3, which gives the  rms noise limits $\sigma T_b$ in mK  scaled to a 1 \kms\ channel for each position and transition (col.~6).
The low-velocity emission line parameters  {\bf and their $1\sigma$ errors} were derived from a Gaussian fit (cols.~3-5).
Figure 3 shows spectra in the four \oh\ transitions 
averaged over the three main directions under study.

\input{Smith_LowVel_OH_tab_v2.tex}

These are very sensitive \oh\ spectra with an 
{\bf rms noise $\sigma T_b \sim 0.7$ mK in a 1 \kms\ channel in the main lines.}
This is comparable to the sensitivity of the low latitudes studies of \citet{Busch2021}.  
In contrast, a more typical modern \oh\ survey, e.g., the SPLASH survey of \citet{Dawson2022}, has a noise level $\approx 15$ mK over a similar velocity interval.



\section{OH in the Smith Cloud}
\label{sec:limits}

\begin{figure}
\label{fig:OHspectra}
	\centering
	\includegraphics[width=0.9\textwidth]{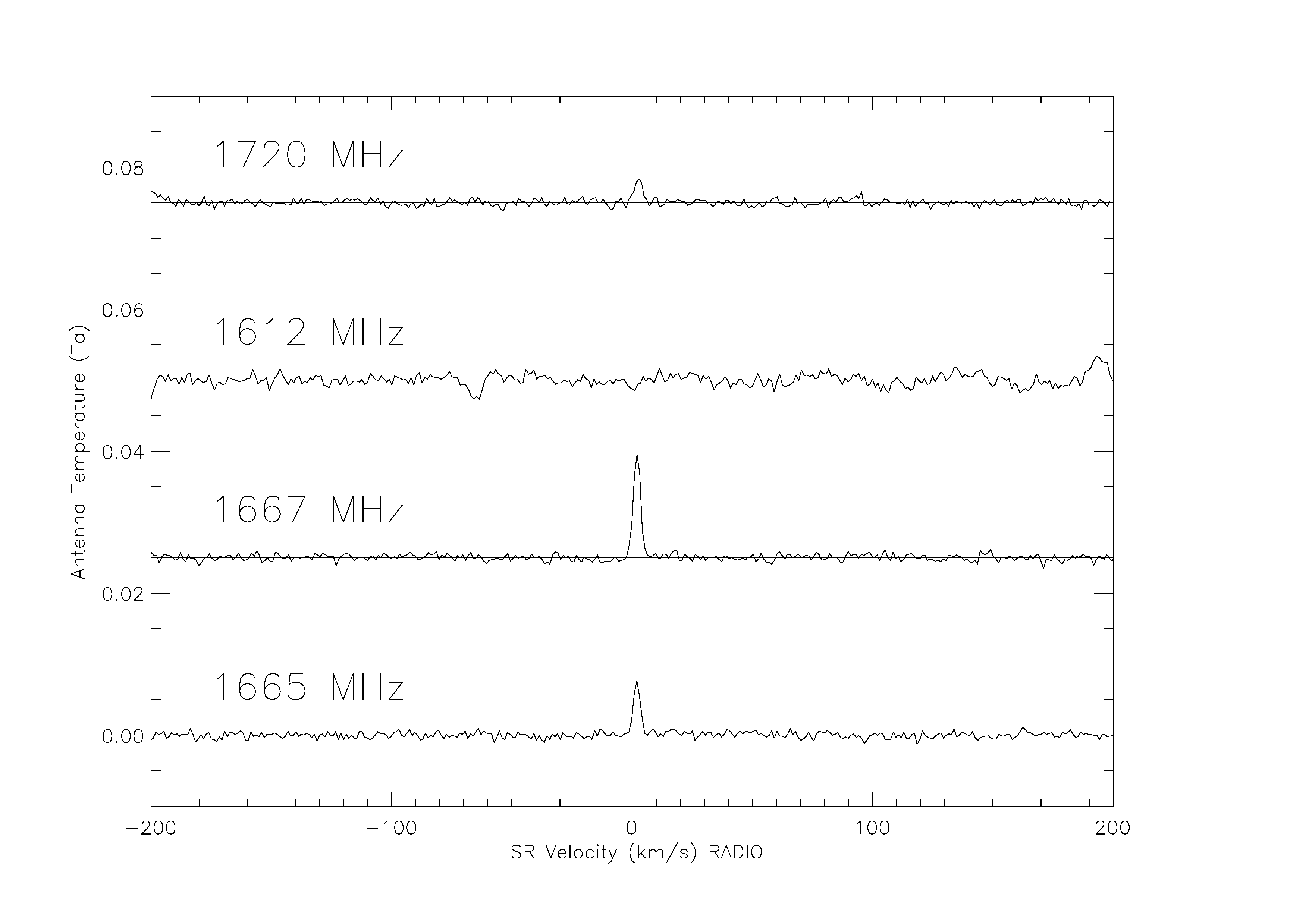}
	\caption{
		GBT \oh\ spectra from the average of the three  positions with the largest \nhi\ that we observed in the Smith Cloud.  
  {\bf Units are antenna temperature. For a spatially extended source $T_a = 0.83 T_b$. }
  The OH lines are offset in 0.025~K increments.  
    Any \oh\ associated with the Smith Cloud
		should appear at $\vlsr \approx +100\ \kms$.
    The only \oh\ detected is at low velocity, associated with the foreground Aquila molecular cloud.
    }
\end{figure}

Table 4 shows \oh\ limits derived for the Smith Cloud.
Assuming Local Thermodynamic Equilibrium (LTE), 
measurement of the main lines of \oh\  can be used to place limits on the column density of the \oh\ molecule  \citep[e.g.,][]{Liszt1996, Barriault2010}.
For the 1667 MHz transition

\begin{equation}
\label{eq:NOH}
 \noh =   2.26\times 10^{14}\  {\Delta T_{ex}} \int T_b(1667) dv  \ \ \cmm  \ \ \ 
 \end{equation}
where $T_b$ is in Kelvins and $v$ is in \kms\ \citep[e.g.,][]{Busch2021}.

The quantity $\Delta T_{ex}$ is related to the \oh\ transition excitation temperature $T_{ex}$ and the continuum background temperature at this frequency $T_{bg}$ through: 

\begin{equation}
\label{eq:Tex}
\label{Tex}
\Delta T_{ex} = {{T_{ex} } \over{(T_{ex} - T_{bg})}}.
\end{equation}

The Smith Cloud is $\approx 3$ kpc from the Galactic Plane so the cosmic microwave background is likely to be the dominant component of the background continuum temperature $T_{bg}$. 
In the Galactic ISM $T_{ex}$ can vary by an order of magnitude (e.g. \citet{Nguyen2018}) and it is not obvious what the appropriate value of $T_{ex}$ should be for molecules in an object as atypical as the Smith Cloud.  
As the Smith Cloud directions lack bright background sources, we expect that $\Delta T_{ex} \geq 1.$ 
In the limit that $T_{ex} \gg T_{bg}$ the quantity $\Delta T_{ex} \to 1$.
\citet{Barriault2010} adopted $\Delta T_{ex} = 1.0$ for the  North Celestial Loop, while \citet{Busch2021} considered values  $4.6 < T_{ex} < 6.1$ K and a range of $T_{bg}$ for their analysis of \oh\  emission on long sitelines through the outer Milky Way disk. 
{\bf While it is conceivable that the excitation temperature of \oh\ in the Smith Cloud is close to $T_{bg}$, allowing for a considerable abundance of \oh\ with little emission in the 18cm lines, the fact that \oh\ emission is detected over a range of \nhi\ in diverse environments throughout the Galaxy (e.g, \citet{Allen2012, Busch2021, Smith2023}), suggests that this circumstance is not common.}
In the absence of any other information we consider $\Delta T_{ex}$  as an unknown, set to unity for convenient scaling.

When there is only a limit on the \oh\ line brightness, we replace the integral in eq.~1 with the noise in a 1 \kms\ channel, $\sigma T_b$, and sum over a velocity interval $\Delta v$ \kms.
\autoref{eq:NOH} can then be written  

\begin{equation}
\label{eq:NOH_limit}
    \noh \leq   7.2\times 10^{11}\  \Delta T_{ex} {\sqrt{\Delta v / 10}} \ \sigma T_b\ \  \cmm  
\end{equation}
for a $1\sigma$ limit where $\sigma T_b$ is in mK and $\Delta v$ is in \kms.

The measurements of $\sigma T_b$ from Table 3 are used to calculate $\sigma \noh$ given in col.~4 of Table 4. 
If any molecular gas in the Smith Cloud is well-mixed with the neutral atomic hydrogen, and turbulence is small compared to thermal motions, \oh\ emission lines should have a FWHM only 0.24 that of \hi.
This is unlikely to be the case as the 21cm lines are broad and often not a single Gaussian.  
We thus adopt $\Delta v = $ FWHM of the 21cm line from Table 1.

Table 4 gives limits on the $\noh/\nhi$ ratios for each observed position in the 1665 MHz and 1667 MHz transitions separately.
A weighted average of the ratios for each transition is given in the row marked "Average" in Table 4.
We combine limits on the two transitions assuming that the 1665 MHz and 1667 MHz lines would be in the LTE ratio of 5:9. 
The final average "combined" limits on the relative abundance of \oh\ in the Smith Cloud are listed in the last row: $\noh /\nhi \leq 3.0 \times 10^{-9}$ at the $5\sigma$ noise level.

Limits on the \oh\ molecule can be used to place limits on the molecular hydrogen column density ${\rm N_{H2}}$ through the relationship 
$\noh /{\rm N_{H2}} = 1.0 \times 10^{-7}$ \citep{Weselak2010,Nguyen2018,Weselak2022} {\bf though we note that this relationship was derived from sightlines in the Galactic disk and might not be correct for low-metallicity gas.}

The quantities in Table 4 are upper limits because no \oh\ emission was detected at the velocity of the Smith Cloud.
They are also, unfortunately, somewhat lower limits because they were derived assuming that  $\Delta T_{ex} = 1$, while it is more likely that $\Delta T_{ex}$ has values of a few.
Nonetheless, for plausible values of $\Delta T_{ex}$ the limits are so low that they do provide information about conditions in the Smith Cloud.

\input{Smith_OH_limits_table_v2.tex}

\section{Dust}
\label{sec:dust}

In the diffuse Galactic ISM there is a good correlation between dust and total  gas, atomic plus molecular.
This correlation can be seen in the reddening parameter E(B-V) and in emission at far-infrared (FIR)  wavelengths \citep{Desert1988,Boulanger1996}.
A more comprehensive analysis of the dust content of the Smith Cloud will accompany the release of the new GBT \hi\ survey; here will discuss evidence for dust in the Smith Cloud along the sightlines searched for \oh.

A recent review and analysis of a large body of data by \citet{Liszt2023} concludes that 1) at $ \nhi \lesssim 10^{21} \cmm$ there is a linear relationship between \nhi\ and the reddening parameter E(B-V) inferred from far infrared (FIR) emission 
\citep{Schlegel1998,Gillmon2006}.
2) Various investigations have found that the ratio of total proton column density to reddening, \nh /E(B-V), is in the range 
$ 6 - 9 \times 10^{21} $   H nuclei \cmm\ ${\rm mag^{-1}}$  
where \nh\ includes contributions from H${_2}$  as well as \hi.

\input{Reddening_table_v7.tex}

To estimate the reddening associated the foreground gas we use the median of the values quoted in \citet{Liszt2023}: 
\begin{equation}
\label{eq:reddening}
{\rm \frac{N_{\rm H}}{E(B-V)} = (7.5\pm1.5) \times 10^{21} \ \  H \ \ nuclei \ \ \cmm\ mag^{-1}}
\end{equation}
 where the errors encompass almost the entire range of values derived by different investigations.
The total \nhi\ towards the Smith Cloud can be divided into a component associated with the Smith Cloud, and foreground component that  contains all \hi\ emission except that associated with the Smith Cloud.
The foreground \hi\  component has a velocity consistent with Galactic rotation; its expected reddening is given in col.~2 of Table 5.

The Smith Cloud is observed behind low-\vlsr\  molecular gas (Table 3) that is likely associated with the Aquila molecular cloud \citep{Dame2001},  
components of which are detected in CO about $10\arcdeg$ from the tip of the Smith Cloud.
The components nearest the Smith Cloud on the sky are at a distance of 400 pc, though other components may be as close as 200 pc \citep{Su2020}.
The main line (1665 MHz, 1667 MHz) \oh\ measurements in Table 3 are used to estimate the contribution to the reddening from this foreground molecular gas using eq.~1 assuming $\Delta T_{ex} = 1$, that $\noh /{\rm N_{H2}} = 1.0 \times 10^{-7}$ \citep{Weselak2010,Nguyen2018,Weselak2022}, and the same relationship between \nh\ and E(B-V) as used for the \hi.
We combine 1667 MHz and 1665 MHz measurements of the foreground \oh\ emission in the LTE ratio of 9:5.
The results, in col.~3 of Table~5, show the explicit dependence on the quantity $\Delta T_{ex}$.

The satellite \oh\ lines at 1612 MHz and 1720 MHz from the foreground molecular gas show clear signs of non-LTE effects and exhibit the "conjugate pair" phenomenon, where one transition (typically the 1612 MHz) is in absorption while the conjugate transition (typically 1720 MHz) is in emission \citep[e.g.,][]{Petzler2020}.
This phenomenon is the norm in the Galactic plane, with 71\% of the directions in the SPLASH survey showing this pattern \citep{Dawson2022}. 
However, it  makes these transitions not useful in estimating column densities of \oh\ 
{\bf without significant effort in radiative transfer modeling, e.g., \citet{Hafner2023}.}

Toward the Smith Cloud  the total radio continuum emission is comprised of the 2.7 K microwave background, and a non-thermal component estimated to be 1.5 K from extrapolation of the \citet{Haslam1981} survey in this direction.
Because the Aquila molecular cloud is so close to the Sun, it is likely that the entire nonthermal component lies in the background, hence $T_{bg} = 2.7 + 1.5 = 4.2$ K.
We adopt the value $T_{ex} = 5.4 $ K, which is in the center of the range of $T_{ex}$ adopted by \citet{Busch2021} and in the range of $T_{ex}$ observed in Galactic molecular clouds \citep{Li2018}.
This results in an excitation parameter $\Delta T_{ex} = 4.5$, which is used to scale the values in col.~3 of Table~5. 
 When added to the reddening from foreground \hi, this produces the estimated total foreground reddening given in col.~4.

An independent estimate of the foreground reddening has been made over the entire sky using far-infrared (FIR) emission calibrated by stellar reddening.  
Column 5 of Table~5 shows such values from \citet{Schlafly2011}. 
A measure of the total line-of-sight reddening through the Galactic disk plus halo can be derived by scaling  the dust opacity at 353 GHz, $E(B-V) =  1.49\times 10^4 \ \ \tau_{353}$, as determined from measurements by the Planck telescope \citep{Planck2014, Planck2016}.
This is given in col.~6.
The difference between reddening associated with foreground gas and the total from the FIR is given in col.~7.
{\it This is the reddening that is not accounted for by known foreground gas and which thus might arise in the Smith Cloud.}
The values near zero indicate that the contribution of the Smith Cloud dust to $\tau_{353}$ GHz must be almost undetectable.
\citet{Shull2024} find that Planck estimates of $E(B-V)$ in low reddening sitelines are $12\%$ higher than the \citet{Schlafly2011} values, but this is not seen toward the Smith Cloud.

The reddening expected from \hi\ in the Smith Cloud, also derived from \autoref{eq:reddening}, is the lower value in col.~8.
The larger value includes any potential reddening from the $5\sigma$ limit on N$_{\rm H2}$ from Table 4.
This analysis is similar to that by \citet{Lenz2016}, \citet{Hayakawa2022}, and others, and shows that all the FIR emissivity can be accounted for from foreground material.  
Given the expected reddening from the Smith Cloud (col.~8) it must have a dust abundance, or a FIR dust emissivity at least a factor of three lower than gas in the Galactic disk.




\section{Constraints on the physical conditions}
\label{sec:phys_cond}
The measured column density of \hi\ and upper limit on \oh\ can be used to derive the constraints on the number density in the medium associated with the Smith cloud. 
We used the model described in \citealt{Balashev2021} to evaluate \oh\ column densities. 
Briefly, this model calculates the abundance of \oh\ in a diffuse cold medium by treating the analytical description of the \hi/$\rm H_2$ transition, ionization balance, and chemical reactions within a hydrogen and oxygen chemical network. 
We assume a plain parallel homogeneous cloud with number density, $n_{\rm tot}$, metallicity, $[Z]$, and fixed temperature, $T=100\,K$. This temperature value is a representative number for the cold diffuse medium, \citep[see, e.g.][]{Balashev2019}. We note that if the cloud corresponds to the warm neutral medium with temperature $T\sim 8000$\,K, then 
 the typical \oh/\hi\ abundance will be $\lesssim 10^{-10}$ \citep[e.g.][]{Balashev2021}. The cloud is exposed by cosmic rays specified by primary ionization rate of hydrogen atom, $\zeta$, in s$^{-1}$ and by UV field denoted by $\chi$ and measured in the Mathis field units  \citep{Mathis1983} from one side. 
The latter mimics the situation of UV flux coming preferentially from the Milky Way. 
Within this model, the \oh\ column density for a given \hi\ column density depends on the four global physical parameters:  metallicity, number density, incident UV field strength, and cosmic ray ionization rate. 
The observed limits on \oh\  can constrain the combination of  physical parameters. 

We followed the Bayesian approach, sampling posterior distribution functions of the parameters using a Python implementation of an affine invariant Monte-Carlo Markov Chain sampler (\citep{Goodman2010} within the {\it emcee} package  \citep{emcee}. 
As was shown in \citet{Balashev2021}, the calculated oxygen-bearing molecular abundances using this model are in agreement with the calculations using the more sophisticated code {\sc Meudon PDR} \citep{LePetit2006}, but the significantly reduced calculation time of our code allow us to efficiently sample the physical parameter space. 
We considered the observational limit on the \oh\ column density as a likelihood function, while keeping the \hi\ column density as model parameter which sets the depth of the cloud\footnote{Note that \oh\ abundance varies within the diffuse cloud, mostly due to changes of $\rm H_2$ abundance \citep[see, e.g.,][]{Hollenbach2012, Balashev2021}. For simplicity, we assume that UV field which defines the \hi/$\rm H_2$ transition falls only on one side of the cloud. We obtain almost the same results if the cloud is exposed to the same UV field from both sides.}. 
To sample the posterior probability function we used the measured \hi\ column density as a prior, along with  Gaussian priors on the logarithm of metallicity, with mean value $[Z]=-0.28 \pm 0.14$ \citep{Fox2016} and on the UV field. 
The latter was constrained to be $\log\chi = -1.2$ using a model of circumgalactic UV field by \cite{Fox2016} and the assumed distance to the Smith cloud. 
We took a conservative dispersion on the UV field to be $0.3$ (dex), taking into account uncertainties of the distance measure and the UV flux model. For number density and $\zeta$ we assumed flat priors in log space, emulating a wide distribution. 

The derived posteriors of the physical conditions for the average combined limit on the \oh\ column density of the Smith cloud are shown in Fig.~4. 
One can see that we can derive an upper limit on the hydrogen number density to be $n_{\rm H}^{\rm tot} < 0.2$\,cm$^{-3}$ $(1\sigma)$, which is consistent with previous estimates by \citet{Fox2016} obtained by photoionization modelling of the ions abundances detected along three sight lines towards quasars which are located at the periphery of the Smith cloud. We note that while our calculations did not consider the thermal balance, this upper limit is near the limiting value for the existence of the cold neutral phase \citep[see e.g.][]{Bialy2019}.
We also derive a lower limit on the ionization parameter consistent with the \citet{Fox2016} value of $\log U > -4$. Since we calculated the $\rm H_2$ abundance within the model, we also can constrain the $\rm H_2$ column density at the level $\log N (\rm H_2) < 16$ and $19$ at the $2\sigma$ and $3\sigma$ confidence interval, respectively. 
Interestingly, the 2d posterior function shown in Fig.~4  suggests that if the number density in Smith cloud is $n \approx 0.1$\,cm$^{-3}$,  we are close to the detection limit of \oh\ and the $\rm H_2$ column density derived from the dust (Table 5). 
Alternatively, if we used an estimate on the number density of $\log n_{\rm H}^{\rm tot}\approx -1.5$ derived by \citet{Fox2016} we should expect the H$_2$ column density to be $N_{\rm H_2} \sim 10^{15}$\,cm$^{-2}$. If we considered each sightline independently (see Table~\ref{tab:OHabundance}), we get slightly less strict upper limits on the number density, but reasonably similar limits on $N_{\rm H_2}$. 
 
\begin{figure}
\label{fig:phys_cond}
	\centering
	\includegraphics[width=0.9\textwidth]{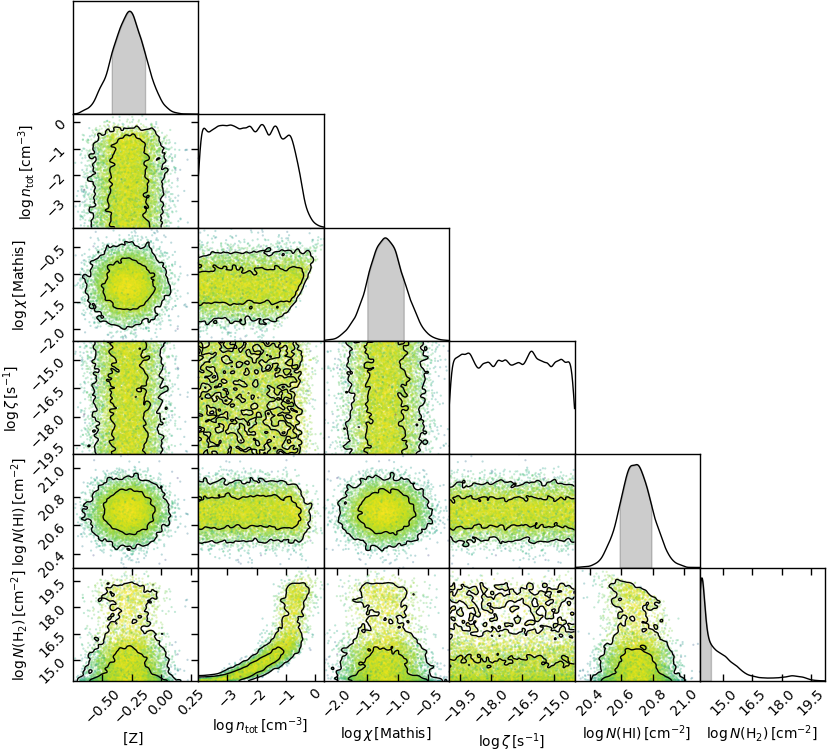}
	\caption{The 2d (off diagonal panels) and 1d (on the diagonal panels) posterior distribution function of the physical parameters and column densities obtained by MCMC sampling \citep{emcee} of the models of \oh/\hi\ abundances. 
 We used average \nhi\ values 
 and upper limits on \oh. 
 The contours on the 2d posterior distribution panels correspond to the 1 and 2$\sigma$ significance intervals, while the gray area marks the 1d posterior distribution corresponding to a 1$\sigma$ significance interval. 
 The 1d posterior function on the metallicity $[Z]$, UV flux ($\chi$) and \hi\ column density coincide with assumed priors from the measurements in this work and other studies, while the upper limits on the number density, ($n_{\rm tot}$) and $\rm H_2$ column density are the main result of the modelling. 
 }
\end{figure}

\section{Conclusions}

The absence of \oh\ emission from the higher \nhi\ regions of the Smith Cloud implies a fractional abundance 
$N_{\rm H_2}/\nhi \ \leq 0.03 \Delta T_{ex}$ at the $5 \sigma$ level.  
This is consistent with calculations of the expected abundance in a low-metallicity diffuse cloud in the Galactic halo (Section 5).

Molecular gas is generally absent from HVCs except for those likely associated with the Magellanic Clouds or the Milky Way's nuclear outflow \citep{Putman2012,DiTeodoro2020,Cashman2021,Tchernyshyov2022}.
Previous searches for evidence of FIR emission from dust in HVCs have been almost uniformly unsuccessful \citep[e.g.,][and references therein]{Lenz2016, Shull2024}, reaching limits more than an order of magnitude lower than typical ISM values.
Recently, however, \citet{Fox2023} have detected evidence of the depletion of refactory elements (Al, Fe, Si) in the HVC Complex C, indicating the presence of dust grains, though the overall metallicity of this HVC is only 0.1 - 0.3 solar \citep[e.g.,][]{Wakker1999,Shull2011}.
In this regard the Smith Cloud seems typical of HVCs even though its deviation velocity, defined as the difference between its space velocity and that expected from Galactic rotation at its location, is not particularly large \citep{Wakker1991}.
Indeed, \citet{Smith1963} considered it most likely that the cloud was a structure associated with the Milky Way disk.
We now know that in contrast to HVCs, Intermediate Velocity Clouds have near-normal metalicities, contain molecules, and are easily detected {\bf in \oh\ and FIR emission  \citep{Putman2012,Lenz2015,Smith2018}.
In this regard IVCs resemble high-latitude diffuse molecular clouds \citep{Magnani1996}. }
{\bf While uncertainties in the excitation temperature of the \oh\ transitions give the possibility that there may be large quantities of OH at undetectable levels in the Smith Cloud, when considering all the evidence we feel that the absence of significant \oh\ emission is not misleading.} 
All the data on the Smith Cloud suggest that it has a low metallicity, a low dust content, a low molecular abundance, and thus a history quite different from that of Milky Way disk gas.
{\bf Further studies of the Cloud's metallicity through UV spectroscopy of absorption line (e.g., \citet{Fox2016}) would make a major contribution our understanding of this unique object.}

\section{Acknowledgments}  The observations were made under GBT proposal codes GBT/13B-274 and GBT/14B-513.  The Green Bank Observatory is a facility of the National Science Foundation, operated under a cooperative agreement by Associated Universities, Inc. 
This research has made use of the NASA/IPAC Infrared Science Archive, which is funded by the National Aeronautics and Space Administration and operated by the California Institute of Technology. 
This research has made use of data, software and/or web tools obtained from the High Energy Astrophysics Science Archive Research Center (HEASARC), a service of the Astrophysics Science Division at NASA/GSFC and of the Smithsonian Astrophysical Observatory's High Energy Astrophysics Division. 
We also used the Planck Legacy Archive and the Argonaut Skymaps.
SB was supported by RSF grant 23-12-00166.
We thank the anonymous referee for useful comments.

\facilities{GBT, IRSA}

\bibliography{SmithCloudOH}
\bibliographystyle{aasjournal}

\end{document}

%% file: Introduction.tex
\label{sec:Introduction}

The Smith Cloud is one of the most prominent Galactic high-velocity clouds \citep{Smith1963}.  
It extends over more than $10{\arcdeg}$ on the sky in neutral hydrogen emission with a highly-organized cometary morphology.    
It contains at least $10^6$ \mo\ of \hi\ and probably an equal mass in ionized gas but no obvious stellar component (\citet{Lockman2008} hereafter L08; \citet{Hill2009, 2015Stark}).  
Its metallicity is about one-half solar with an average $[Z]=-0.28 \pm 0.14$, and it is draped in a magnetic field  \citep{Hill2009,Fox2016,Betti2019}.
 
 There are several proposals for the origin of the Smith Cloud.
 It could be the  tidal remnant of an accreting dwarf galaxy; the baryonic component of a dark matter sub-halo; the product of a galactic fountain driven by multiple supernovae in the Milk Way disk \citep{Bland-Hawthorn1998, 2009Nichols, 2014Nichols, 2016Galyardt, 2017Marasco}. 
 If the Smith Cloud is indeed a tidal remnant, or is the product of a Galactic fountain, it may contain a molecular phase in addition to its \hi\ and H$^+$.
 
 At the $9.1\arcmin$ angular resolution of the GBT 
the 21cm neutral hydrogen emission from the Smith Cloud reaches a maximum brightness temperature $T_b \approx 15$ K, and a maximum column density $\nhi \approx 6 \times 10^{20}\  \cmm$.  
This value of \nhi\ is several times greater than the atomic-to-molecular transition observed at high Galactic latitudes \citep{Gillmon2006} and is comparable to high-latitude Galactic sightlines that have a molecular hydrogen fraction $\gtrsim 10\%$ \citep{Liszt2023,Shull2024}. 
Diffuse interstellar clouds, and even some intermediate velocity clouds at this \hi\ column density have a molecular component that is detectable in the 18 cm radio lines of \oh\ \citep{Barriault2010,Smith2018,Tang2021}. 
Moreover, in a study of a sightlines through the outer Milky Way,  \citet{Busch2021} detect \oh\ emission in the 1667 MHz line at intensities $T_b(1667) \gtrsim 4$ mK at all velocities where the \hi\ emission is $T_b(\rm 21cm) \gtrsim 10$ K, an \hi\ brightness temperature in the range of that seen in the Smith Cloud. 

Because the relative abundance ratios ${\rm OH/HI}$, and ${\rm OH/H_2}$  are  sensitive to  physical conditions in diffuse interstellar gas \citep[e.g.,][]{Balashev2021}, 
we used the GBT to search for emission from the 18cm transitions of \oh\ at five positions in the Cloud.
In addition, we searched for \oh\ over a 1 sq-degree field at the tip of the Cloud, where there is evidence of CO emission from the Planck survey \citep{Planck2014}, {\bf though that survey detects only total CO and cannot distinguish kinematically between 
 foreground gas and the Smith Cloud.}
Three directions where $T_B(\mbox{\hi}) > 14$ K and $\nhi \geq 4.5 \times 10^{20}$ \cmm\ were observed to a sensitivity level sufficient to detect the 1667 MHz line from an equivalent Galactic interstellar cloud, or from the  interstellar gas detected in \oh\ by \citet{Busch2021} on long paths through the outer Galactic disk.  
Although we will focus on the data from these three directions as they give the most sensitive limits on \oh\ in the Smith Cloud,  our conclusions apply to the other fields as well but with less strict limits.

\citet{Tang2021} give a recent summary of expectations for \oh\ emission.
There are four 18 cm transitions of {\bf ground state} \oh, two main lines at 1665 and 1667 MHz, and satellite lines at 1612 and 1720 MHz. 
In Local Thermodynamic Equilibrium (LTE) the 1665 and 1667 MHz lines should be 5 and 9 times brighter than the satellite lines, respectively.
Local effects arising from the intensity of the infrared radiation field commonly cause deviations from the expected line ratios.
It is not clear what to expect for \oh\ excitation conditions in the Smith Cloud, as it lies in the lower halo where local sources of infrared radiation should be rare, but where the Cloud is impacted by diffuse ionizing radiation  escaping from the Galactic disk \citep{Bland-Hawthorn1998,Bland-Hawthorn2002,Putman2003,Fox2016}.

 For this paper we adopt the basic properties of the Smith Cloud from L08: the densest parts of the Cloud are  $12.4\pm1.3$ kpc from the Sun, at  $z = -2.9\pm0.3$ kpc from the Galactic plane; its extent in \hi\  is $>3$ kpc.  The Cloud lies toward the inner galaxy only $\approx 15\arcdeg$ from the Galactic plane and is viewed through a significant foreground of unrelated gas and dust.  
 
 In Section 2 we describe the observations of \hi\ and \oh. 
 Section 3 describes the conversion of the observations into limits on N(OH) and related quantities.
 The question of dust in the Smith Cloud is examined in Section 4.
 The implications of these results on models for the Smith Cloud are discussed in Section 5, and 
 Section 6 gives a brief summary.
 

%% file: HI_observations.tex
\label{sec:HIObservations}
For this work we use 21cm \hi\ observations from the GBT made as part of a new survey of the Smith Cloud.
These data have improvements in calibration, velocity coverage, and velocity resolution compared to the data presented in L08.
Data were acquired by in-band frequency switching  and reduced to include all $|\vlsr| \lesssim 400\ \kms$  at  resolution of 1.03 \kms. 
Spectra were calibrated, corrected for stray radiation, and converted to brightness temperature using the method described in \citet{Boothroyd2011}.
A low-order polynomial was fit to emission-free regions of the spectra with a resultant rms noise in brightness temperature of 67 mK in a 1 \kms\ channel.

The three principle positions searched for \oh\ are marked on a WISE $22\mu$ image along with contours of \nhi\ in Figure 1, and the measured \hi\ spectra  are given in Figure 2.
The Smith Cloud \hi\ emission is not faint; it can constitute $\gtrsim 30\%$ of the total \nhi\ in these directions.

\begin{figure}
    \centering
    \gridline{\fig{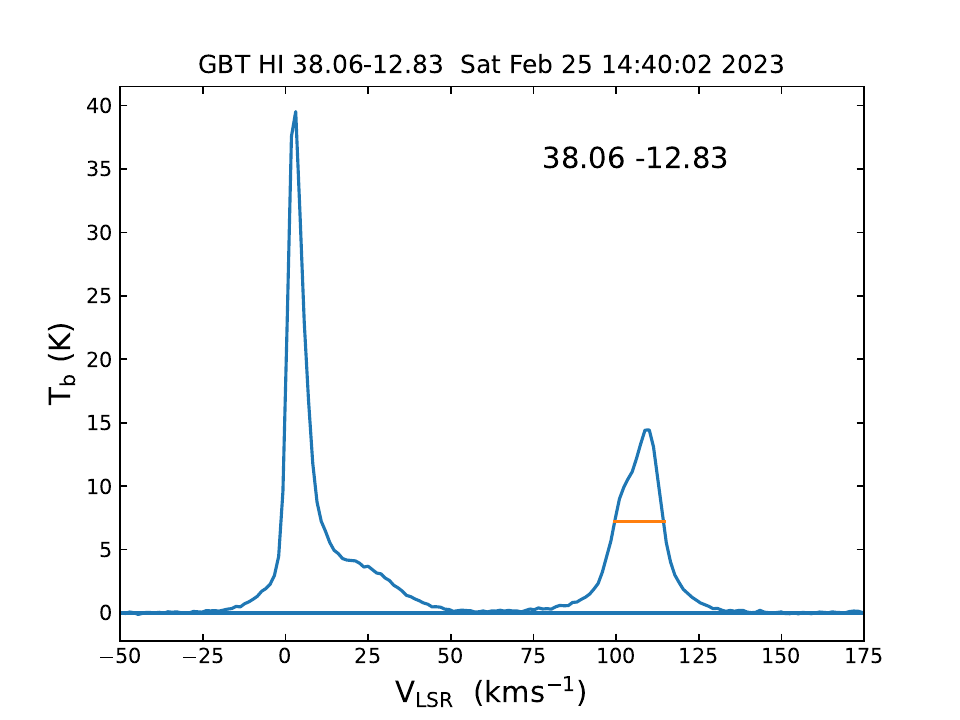}{0.3\textwidth}{}
              \fig{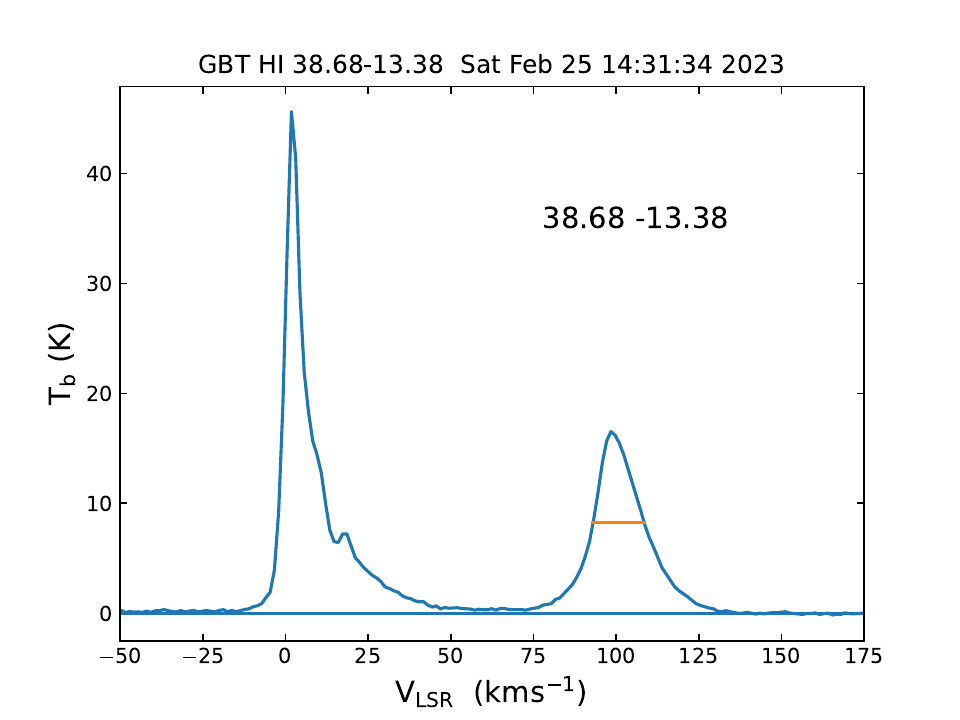}{0.3\textwidth}{}
              \fig{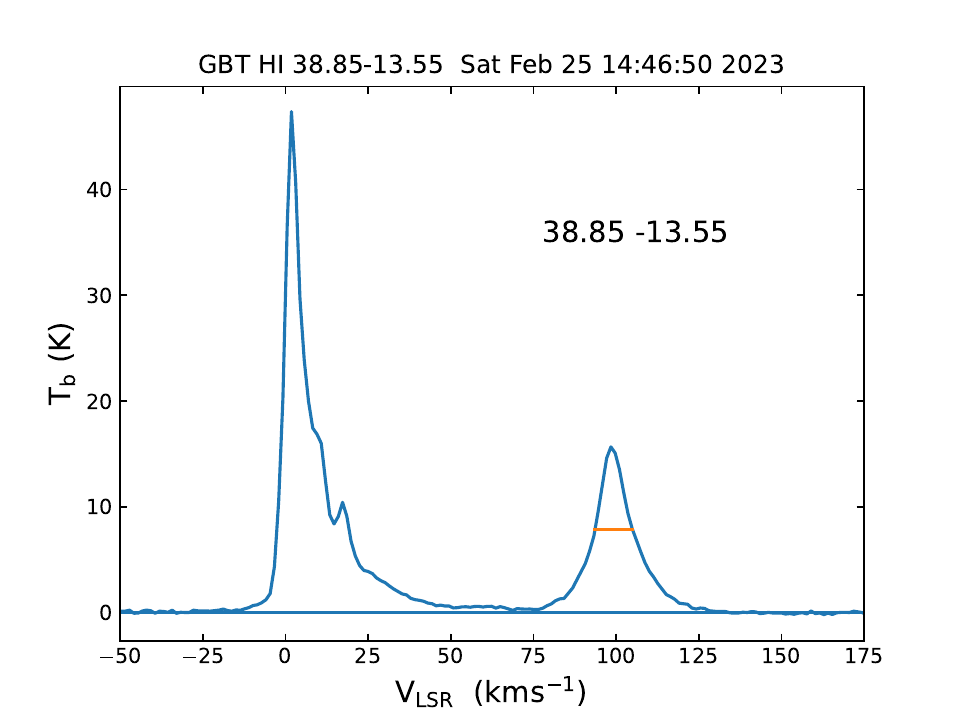}{0.3\textwidth}{}}
    \caption{21cm \hi\ spectra toward the  Smith Cloud directions that give the most strict limits on \oh. 
    The Smith Cloud emission occurs near 100 \kms; its FWHM is indicated by a horizontal line.
    }
    \label{fig:HI_spectra}
\end{figure}

The \hi\ properties  are summarized in Table 2, where columns 1 and 2 give the Galactic coordinates 
{\bf and we divide the \hi\ into foreground gas at $\vlsr < 70$ \kms, and Smith Cloud emission which peaks around 100 \kms.}  
Values of \nhi\ were calculated from the usual relationships \citep{DickeyLockman1990} with an assumed \hi\ excitation temperature (spin temperature) of 100 K.
The correction for opacity  increases \nhi\ by only $5\%$ above the optically thin value.
The \hi\ lines from the Smith Cloud do not have a simple Gaussian shape.
The quantity $T_{pk}$ in col.~7 refers to the maximum brightness temperature in the line, and the FWHM (col.~8) is the velocity width measured at half the peak brightness.

The observed  positions lie along a bright, fairly narrow ridge in the \hi\ emission that has a FWHM $\sim 0\fdg3$.
At the adopted distance of the Smith Cloud this corresponds to a transverse linear size of 65 pc.
Assuming cylindrical symmetry, we use this value to convert \nhi\ to an average volume density,{\bf  $<n_{HI}>$,} that is shown in col.~10 of Table 2.

%% file: Smith_HI_tab.tex
\begin{deluxetable}{cccccccccc}
\tabletypesize{\footnotesize}
\tablecolumns{10}
	\tablecaption{HI in Directions Searched for OH  }
\label{tab:HIdata}
\tablehead{\colhead{$\ell$}  & \colhead{$b$} & \colhead{$\rm N_{HI} (for)$} & \colhead{$\rm T_{pk} (for)$} & 
  \colhead{\rm \vlsr (for)} & \colhead{$\rm N_{HI} {\rm (Smith)}$} &  \colhead{$\rm T_{pk} (Smith)$} &
\colhead{FWHM (Smith)} & \colhead{\rm \vlsr (Smith)} &
\colhead{\rm $<n_{HI}>$ (Smith)} \\ 
\colhead{($\arcdeg$)} & \colhead{($\arcdeg$)} & \colhead{($10^{20}$ \cmm)} & \colhead{(K)} &  \colhead{(\kms)} &
\colhead{($10^{20}$ \cmm)} &  \colhead{(K)} & \colhead{(\kms)} &  \colhead{(\kms)} &
\colhead{$\rm (cm^{-3}$)} \\
\colhead{(1)} &  \colhead{(2)} & \colhead{(3)} & \colhead{(4)}  & \colhead{(5)} & \colhead{(6)} & \colhead{(7)} & \colhead{(8)} & \colhead{(9)} & \colhead{(10)}
}
\startdata
	\hline\hline\noalign{\vspace{5pt}}	
			\noalign{\smallskip}
38.06 & -12.83 & 8.2 & 39.5 & 3.0  & 4.6 & 14.4 & 15.0 & 109.0 & 2.3 \\
38.68 & -13.38 & 10.2 & 45.6 & 1.9 & 5.9 & 16.5 & 15.5 & 100.2 & 2.9 \\
38.85 & -13.55 & 11.2 & 47.4 & 1.9 & 4.5 & 15.7 & 11.4 & 99.0 & 2.2 \\ 
\noalign{\vspace{5pt}}
\enddata
\tablecomments{
The hydrogen column density is calculated for a spin temperature of $100$ K.  {\bf The $1\sigma$ uncertainty in 
$T_{pk}$ is $\approx 0.07$ K.  Velocities are for the brightest line emission in the foreground (for) or Smith Cloud component.} }
\end{deluxetable}

%% file: Smith_LowVel_OH_tab_v2.tex
\begin{deluxetable}{cccccccc}
	\tablecolumns{8}
	\tablecaption{Low Velocity OH Detections} 
 \label{tab:OHobservations}
\tablehead{\colhead{$\ell, b$} & \colhead{Transition} & \colhead{$T_{pk}$} &
\colhead{FWHM} & \colhead{\vlsr}  & \colhead{$\rm \sigma T_b$} & \colhead{$\rm W(OH)$} & \colhead{$N_{OH}$}  
\\
\colhead{} & \colhead{(MHz)} & \colhead{(mK)} &   \colhead{($\rm km ~s^{-1}$)} & \colhead{($\rm km ~s^{-1}$)}  & \colhead {(mK)}  & \colhead{(mK-\kms)} & \colhead{($10^{13} \cmm$)} \\
\colhead{(1)} &  \colhead{(2)} & \colhead{(3)} & \colhead{(4)}  & \colhead{(5)} & \colhead{(6)} & \colhead{(7)} 
& \colhead{(8)} \\
}
\startdata
	\hline\hline\noalign{\vspace{5pt}}	
			\noalign{\smallskip}
38.06-12.83 & 1612  & $-2\pm1$     & $2\pm2$       & $1.8\pm0.6 $   & 0.88 &  & \\ 
            & 1665  & $12.9\pm0.6$ & $2.5\pm0.1$   & $2.76\pm0.06$ &  0.67  & 34.4 & 1.4\\
            & 1667 & $27.4\pm0.7$  & $2.33\pm0.07$ & $2.82\pm0.03$  &  0.59 & 68.0 & 1.5 \\ 
            & 1720 & $6.8\pm0.6$   & $2.5\pm0.2$   & $3.7\pm0.1$    &  0.61 \\
38.68-13.38 & 1612 & $-4\pm11$     & $1\pm4$       & $1.8\pm0.2$    &  1.04 \\
            &  1665 & $4.3\pm0.7$  & $2.9\pm0.5$   & $1.4\pm0.2$   &  0.70  & 13.3 & 0.5\\
            &  1667 & $11.6\pm0.7$ & $2.5\pm0.2$   & $1.62\pm0.08$ &  0.82  & 30.9 & 0.7\\
            &  1720 & $2.5\pm0.5$  & $4\pm1$       & $0.3\pm0.5$   &  0.64  \\
38.85-13.55 & 1612  & $-3\pm1$     & $3\pm1$       & $0.1\pm0.4$   &  1.21  \\
            &  1665 & $11.3\pm0.8$ & $2.4\pm0.2$   & $1.36\pm0.08$ &  0.67  & 28.9 & 1.2\\
            &  1667 & $18.5\pm0.7$ & $2.7\pm0.1$   & $1.16\pm0.05$ &  0.71  & 53.2 & 1.2\\
            &  1720 & $4.9\pm0.6$ & $2.4\pm0.4$    & $0.6\pm0.2$   &  0.68  \\
\noalign{\vspace{5pt}}
\enddata
\tablecomments{Quantities $T_{pk}$, FWHM and ${\rm V_{LSR}}$ are from a Gaussian fit to the \textbf{ \oh\  lines, associated with the Aquila molecular cloud. 
 {\rm $N_{OH}$} is derived from \autoref{eq:NOH} using $\Delta T_{ex} = 1$,  scaled by 9/5 for the 1665 MHz line.
} 
}
\end{deluxetable}

%% file: Smith_OH_limits_table_v2.tex
\begin{deluxetable}{ccccccc}
	\tablecolumns{7}
	\tablecaption{Limits on \oh\ in the Smith Cloud }
\label{tab:OHabundance}
\tablehead{\colhead{Position} & \colhead{Transition} & \colhead{{\rm $\sigma W(\rm OH)$}} &
\colhead{$\sigma \noh$} & \colhead{$\noh/\nhi$} & 
\colhead{{\rm $N(\rm H_2)$}} & \colhead{{\rm $N(\rm H_2)/\nhi$}} \\
\colhead{($\ell, b$)} & \colhead{(MHz)} & \colhead{(mK-\kms)} &  \colhead{($10^{11} \cmm$)}   & \colhead{$(\rm 5\sigma \ limit)$}
& \colhead{($5\sigma$  limit)} & \colhead{($\rm 5\sigma \ limit$)} \\
\colhead{(1)} &  \colhead{(2)} & \colhead{(3)} & \colhead{(4)}  & \colhead{(5)} & \colhead{(6)} & \colhead{(7)} \\
}
\startdata
	\hline\hline\noalign{\vspace{5pt}}	
			\noalign{\smallskip}
$38\fdg06 -12\fdg83$ & 1665 & 2.6 & 10.6 & $1.1$e-8 & $5.3$e19  & $1.1$e-1 \\ 
                    & 1667 & 2.3 &  5.2 & $5.6$e-9 & $2.6$e19  & $5.6$e-2 \\
$38\fdg68 -13\fdg38$ & 1665 & 3.0 & 12.2 & $1.0$e-8 & $6.1$e19  & $1.0$e-1 \\ 
                    & 1667 & 2.8 &  6.3 & $5.4$e-9 & $3.2$e19  & $5.4$e-2 \\ 
$38\fdg85 -13\fdg55$ & 1665 & 2.4 &  9.6 & $1.1$e-8 & $4.8$e19  & $1.1$e-1 \\ 
                    & 1667 & 2.8 &  6.3 & $7.0$e-9 & $3.1$e19  & $7.0$e-2 \\ 
 \hline\noalign{\vspace{3pt}}
Average            & 1665  & 1.5 &  6.1 & $6.2$e-9 & $3.1$e19 & $6.1$e-2  \\
                   & 1667  & 1.5 &  3.4 & $3.4$e-9 & $1.7$e19 & $3.4$e-2  \\
\hline\noalign{\vspace{3pt}}
Average         & Combined &    &  3.0 &  $3.0$e-9 & $1.5$e19 & $3.0$e-2 \\
\noalign{\vspace{5pt}}
\enddata
\tablecomments{Limits are for $\Delta T_{ex} = 1.0$.  
Average ``combined" limits assume that the 1665 MHz and 1667 MHz transitions are in the LTE ratio.
}
\end{deluxetable}

%% file: Reddening_table_v7.tex
\begin{deluxetable}{cccccccc}
   \tablecolumns{8}
   \tablecaption{Reddening E(B-V) in Directions Searched for OH}
\label{tab:Reddeningdata}
\tablehead{
\colhead{Position} &  
\colhead{ \hi\tablenotemark{a}} & 
\colhead{${\rm H_2}$\tablenotemark{b}}  & 
\colhead{Total foreground\tablenotemark{c}} &
\colhead{ FIR disk\tablenotemark{d}}  &
\colhead{ $\tau_{353}$\tablenotemark{e}} & 
\colhead{Difference\tablenotemark{f}} & 
\colhead{Smith\tablenotemark{g}}    \\
\colhead{}  & 
\colhead{(mag)} &
\colhead{(mag)} &
\colhead{(mag)} &
\colhead{(mag)} &
\colhead{(mag)} & 
\colhead{(mag)} & 
\colhead{(mag)} \\
\colhead{(1)} &  \colhead{(2)} & \colhead{(3)} & \colhead{(4)}  & \colhead{(5)} 
& \colhead{(6)} & \colhead{(7)} & \colhead{(8)} 
}
\startdata
	\hline\hline\noalign{\vspace{5pt}}	
			\noalign{\smallskip}
38.06--12.83 & $0.11 \pm 0.02$ & $0.020\Delta T_{ex}$ & $0.20 \pm 0.03$ & 0.19 &0.20 &$-0.00\pm 0.03$ & 0.06 -- 0.07 \\
38.68--13.38 & $0.14 \pm 0.03$ & $0.008\Delta T_{ex}$ & $0.18 \pm 0.03$ & 0.18 &0.18 &$0.00\pm 0.03$ & 0.08 -- 0.09  \\ 
38.85--13.55 & $0.15 \pm 0.03$ & $0.016\Delta T_{ex}$ & $0.22 \pm 0.03$ & 0.24 &0.24 &$0.02 \pm 0.03$ & 0.06 -- 0.07  \\
\noalign{\vspace{5pt}}
\enddata
\tablenotetext{a}{Expected reddening from foreground \hi, i.e.,  omitting the Smith Cloud.} 
\tablenotetext{b}{Reddening from foreground molecular gas detected in low-velocity \oh\ emission.}
\tablenotetext{c}{Total foreground reddening from \hi\ and \oh\ for $\Delta T_{ex} = 4.5$.} 
\tablenotetext{d}{Foreground reddening from \citet{Schlafly2011}}
\tablenotetext{e}{Total reddening disk plus halo from FIR emission.}
\tablenotetext{f}{The difference col.~6 - col.~4 that could arise in the Smith Cloud.}
\tablenotetext{g}{Expected reddening from the Smith Cloud if its dust emission is identical to disk gas.} 
\end{deluxetable}